# Interdependent-Network Criticality without a Second Network: Cascading Collapse of the Joule-Coupled Insulator–Metal Transition in $VO_2$

Ouriel Gotesdyner[1], Michal Wasserman[1,2], Hodaya Kaster[1,2], Yosi Abulafia[1,2], Avital Fried[1,2], Bnaya Gross[3], Shlomo Havlin[1], and Amos Sharoni[1,2,†]

[1]Department of Physics, Bar Ilan University, Ramat-Gan, 5290002, Israel.

[2]Institute of Nanotechnology & Advanced Materials, Bar Ilan University, Ramat-Gan, 5290002, Israel.

[3]Department of Mathematics, Bar Ilan University, Ramat-Gan, 5290002, Israel.

[†]Corresponding author.

**Abstract**

When driven by a large enough electric field or current, the insulating state of vanadium dioxide ($VO_2$) typically collapses abruptly into the metallic state (the insulator–metal transition, IMT) following a brief incubation delay. These incubation delays, typically lasting microseconds or less, have been analyzed predominantly through macroscopic electro-thermal modeling, providing little statistical insight into the transition dynamics. Here, we investigate the electrically driven IMT in a two-dimensional $VO_2$ lattice and demonstrate that the material behaves as a complex, interdependent network despite being a single structural entity. Because metallic domains dissipate significantly more Joule heat than insulating regions, each switching event raises the Joule heating of the network, resulting in network dependency and triggering a cascading chain reaction. Combining time-resolved resistance measurements with a Joule-coupled resistor-network model and interdependent-network theory, we observe an abrupt (mixed-order) transition preceded by an anomalously long incubation plateau lasting thousands of seconds. During this metastable plateau, an effective branching factor of the switching activity approaches unity at criticality, and the plateau lifetime diverges with exponent $\zeta = 1/2$; the order parameter scales with exponent $\beta = 1/2$ . These signatures coincide with those predicted for interdependent networks, indicating that dissipative Joule coupling alone can drive mixed-order catastrophic cascades in a single-network correlated-oxide system.

## I. INTRODUCTION

The abrupt collapse of a macroscopic state under a smoothly varied control parameter is a hallmark of strongly driven condensed-matter systems. How a continuous drive produces a discontinuous response, and whether such collapses share a common organizing principle across otherwise unrelated systems, is a recurring question in statistical physics. A quintessential example is the insulator-metal transition (IMT) of vanadium dioxide ($VO_2$), whose resistance drops by four orders of magnitude near 340 K [1]. In granular thin films the thermally driven transition proceeds as a gradual, random percolation of metallic domains resulting in a continuous transition [2–5]. The same transition can be driven electrically, which is the basis of various applications in neuromorphic computing and resistive-switching memory devices [6–11]. Driven by a sufficiently large field the collapse becomes abrupt, but only after an “incubation” delay, reported from below microseconds up to milliseconds and analyzed almost exclusively through detailed electro-thermal (Joule-heating) modeling [12–17]. Such models, including resistor-network models with local Joule heating [12,18–20], reproduce the switching, yet do not assign the collapse to any statistical or thermodynamic class, leaving its microscopic organization and its true temporal nature unresolved.

The characteristics of the electrically driven transition are set by local electrothermal feedback. Under a constant current, switching an insulating domain to the metallic state lowers its resistance and hence the Joule power it dissipates ($I^2R$). Under a constant voltage the dynamic is reversed: a local switch to the metallic phase raises the dissipated power ($V^2/R$), and the excess heat diffuses to neighboring domains, elevating their effective temperature and driving them to switch in turn [21]. This local dissipative coupling is in essence a (spatially distributed) dependency that can drive the system into a runaway avalanche [14,15]. The transition proceeds neither uniformly nor instantaneously: metallic domains nucleate and grow, switching advances through discrete avalanches, and close to threshold the system exhibits pronounced incubation delays and spatial inhomogeneity [3,9,22,23].

As we show, the resulting voltage-driven collapse is not instantaneous but can be preceded by a pronounced, remarkably slow cascade that can extend beyond $10^4$ seconds, orders of magnitude longer than previously reported. To identify the statistical process behind this delayed collapse we turn to interdependent networks, distinct networks coupled so that the failure of an element in one triggers failures in the other, which undergo abrupt, mixed-order transitions qualitatively unlike ordinary percolation [24]. The order of the transition is set by the strength of the coupling [25], and for spatially embedded networks the collapse becomes discontinuous for any nonzero density of dependency links [26,27]. These transitions carry a distinctive fingerprint: a discontinuous jump appearing only above a coupling threshold; a long-lived metastable plateau, or incubation, preceding the collapse; microscopic cascades whose branching factor approaches unity at the critical point [28,29]; and characteristic critical exponents. This behavior has recently been realized in physical systems of interdependent superconductors coupled by heat [30], self-coupled lasers [31], and proposed for thermally coupled ferromagnets [32], each identifying a positive-feedback channel that acts as a dependency link, now understood within a common cascade universality [33,34]. In almost all the realizations, however, the dependency was engineered between two coupled networks.

Here we recognize that a voltage-driven $VO_2$ film supplies this dependency channel intrinsically, within a single network, without requiring a second network [35–37]. Patterned into a two-dimensional lattice, it becomes a spatial resistor network whose links switch between insulating and metallic states. Joule heating is then the dependency link that thermally intertwines the domains, and the positive feedback that previous realizations engineered between two networks is already present within one. Indeed, cascading first-order collapse has been shown to arise in a single spatially embedded network carrying both connectivity and dependency links [38], and in single systems governed by a self-coupling feedback [31,39], which can naturally apply to Joule-coupled domains in $VO_2$ [10,18,19]. This poses a sharp question: is a single network, endowed only with dissipative (Joule) coupling, sufficient to reproduce the full mixed-order, cascading phenomenology, or is genuine interdependence between two distinct networks essential?

In this Letter we address this question by mapping the electrically driven IMT with time-resolved resistance measurements on a patterned-$VO_2$ lattice together with a Joule-coupled resistor-network model. Both exhibit the same voltage-controlled crossover from a continuous, second-order, temperature-driven percolation transition to a mixed-order, coupling-driven cascading transition. Approaching the critical point, the system develops a long-lived metastable incubation plateau whose duration diverges following a scaling law with exponent $\zeta = 1/2$ before collapsing in a single avalanche. Resolving the cascade in time, we find that the branching factor approaches unity during the plateau, the hallmark of critical, epidemic-like spreading. From slow temperature-dependent resistance measurements near the abrupt transition, we extract an order-parameter exponent β = 1/2. These exponents match those predicted for interdependent networks [40], establishing that Joule heating alone is sufficient to drive interdependent, mixed-order cascades in a single correlated-oxide network, and extending interdependent-network physics beyond the two-network paradigm to a broad class of single, spatially coupled physical systems.

## II. RESULTS

High-quality 65-nm-thick $VO_2$ thin films were deposited on $r$-cut sapphire substrates via reactive rf magnetron sputtering from a 2" $V_2O_3$ target (99.9%, ACI Alloys, Inc.) [41]. A two-dimensional resistor network with a 500 nm wire width, 1 $\mu$m pitch, and total footprint of $200 \times 200$ $\mu\text{m}^2$ was patterned using electron-beam lithography followed by reactive ion etching. Metallic electrodes (10 nm Cr/80 nm Au) were subsequently defined in a second lithography step and deposited via electron-beam evaporation [Fig. 1(a) inset is a cartoon of the device; Fig. 1(b) is an SEM image of part of the device]. Full fabrication details are provided in Supplemental Material (SM) Sec. S1.

Electrical measurements were conducted in a home-made temperature control system with temperature stability better than 1 mK rms (below 6 mK peak-to-peak; SM Sec. S2). Identical results were obtained using both two-probe and four-probe configurations. A critical experimental challenge is isolating local network temperature dynamics while maintaining the heat bath at a constant ambient temperature. Achieving precise thermal control requires strong coupling to the heat bath, yet we wish to avoid excessive external thermal damping that could suppress intrinsic local heating. Balancing these constraints alongside a strict measurement protocol ensures that intrinsic percolation dynamics and latent heat contributions from the phase transition are accurately captured without negative feedback artifacts (SM Sec. S2).

To numerically follow the IMT we used computer simulations to model a 2D lattice composed of nodes and edges and connected to a constant voltage source as shown in Fig. 1(a). Each of the edges is assigned its own critical temperature at which it switches between a phase of high resistance (insulator-like) and a phase of low resistance (metal-like). The distribution of critical temperatures is assigned to the different edges according to a normal distribution around a certain average $T_0$ which depends on the material (for $VO_2$, $T_0 = 340K$) and a standard deviation $\sigma$ which defines the width of the transition ($\sigma \approx 1K$ ) and serves as the disorder of the system.

We impose different values of the external temperature $T$ and observe their effect on the system in the following iterative manner: For each external temperature $T$ we determine the state of each link (high resistance if the temperature is below the link's critical temperature and low resistance if it is above it). Then, we solve the Kirchhoff equations to determine the current and resistance of each link in the lattice and determine the heat generated by each link according to Joule's law

$$Q_i(t) = \kappa \cdot I_i(t)^2 R(t)$$

Where $Q_i(t)$ is the generated heat dissipating from the link i at time t, $I_i(t)$ is the current flow in link i at time t, $R_i(t)$ is the resistance of link i at time t, and $\kappa$ is the dissipation coefficient, which converts dissipated power into a temperature rise (i.e., it plays the role of a thermal resistance to the bath). We then calculate the heat generation of the entire system to determine the effective temperature of the system in the following time step according to

$$T_{eff}(t+1) = T + \frac{1}{N}\sum_{i=1}^{N} Q_i(t) \ .$$

We repeat the process of updating the state of the different links and recalculating the current, resistance and new effective temperature until the system reaches a stable state, before we repeat this process for the next value of external temperature $T$ . We calculate the overall resistance $R(T)$ of the network for each value of $T$ as our order parameter.

It is important to note that in our system a link in a metal state will emit more heat than a link in an insulator-like state, adding a cascading dynamic to the process of externally heating the system due to its internal heat emission. Thus, each link is coupled to its peers not only through the electrical connectivity governed by Kirchhoff's equations, but also through a dependency on their state, via the heat they dissipate. Note that in this minimal model the dependency is global: all links share the same effective temperature $T_{eff}$, and since $\Sigma_i I_i^2 R_i = V^2 G$ for a voltage-biased network, the Joule feedback depends on the network conductance G, whereas the lattice enters through the percolative, Kirchhoff-governed dependence of G on the link states.

We first present the voltage-driven crossover from a continuous to an abrupt IMT. Figure 1(a) shows the resistance versus temperature (R-T) measured at a ramp rate of 0.1 K/min under an applied voltage of 12V. The ramp-rate dependence was characterized over a range of sweep rates (SM Sec. S2), with 0.1 K/min selected as it had no significant effect on the R–T curve. Abrupt resistance drops occur during both heating and cooling (see Fig. 1(a)); hereafter, we focus on the transition behavior along the heating branch.

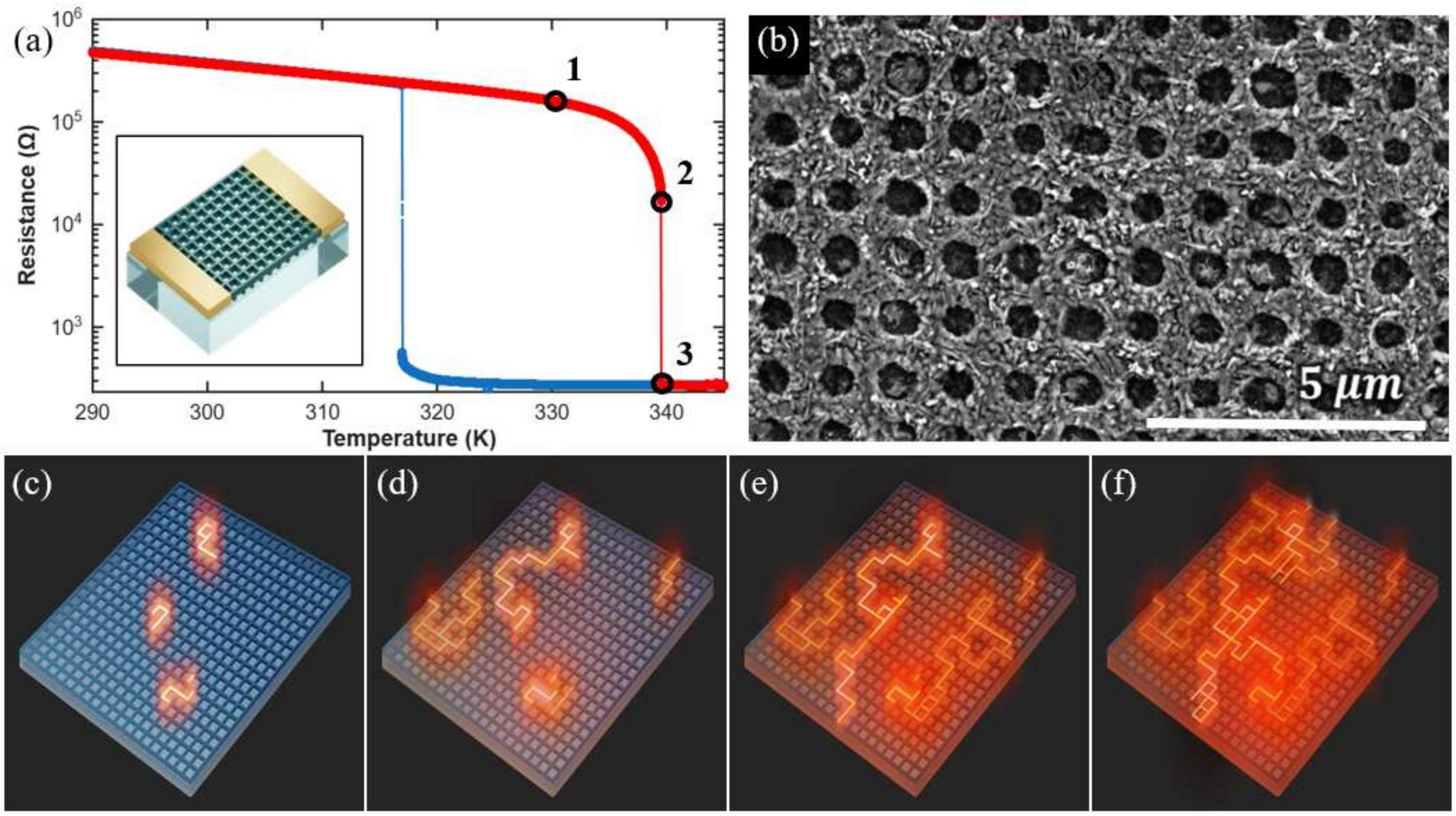

FIG. 1. (a) Resistance versus temperature ($R$-$T$) hysteresis loop of the $VO_2$ network device during heating (red) and cooling (blue). Numbered points correspond to transition stages illustrated in (c)–(f). Inset: Schematic of the network device. (b) SEM image of the device, showing $VO_2$ nanowires (~500 nm wide) separated by 500×500 $nm^2$ gaps. (c)–(f) Schematic of the cascading phase transition mechanism. (c) At $T < T_C$ under applied bias V, local regions with lower transition temperatures act as metallic nucleation sites, (point 1 in (a)). (d) In addition to the temperature increase, localized Joule heating ($V^2/R$) at sites diffuses through the substrate, self-warming adjacent areas and nucleating further transitions, (point 2 in (a)). (e) Above a critical voltage $V_C$, thermal feedback drives regions with higher local $T_C$ metallic, triggering a percolation cascade that establishes a continuous conductive path across the network, (between points 2 and 3 in (a)). (f) Post-percolation self-heating recruits further high-$T_C$ regions, expanding the metallic domain throughout the network (point 3 in (a)).

Figure 2(a) presents $R$–$T$ characteristics of the same device across various applied voltages, following initial cooling to 290 K and subsequent heating to 350 K at 0.1 K/min. At low voltages (e.g., $V = 0.5$ V), the IMT remains continuous. Above a critical threshold ($V_c \approx 5.5$ V), the transition becomes abrupt, with the onset temperature of the resistance drop, $T_c(V)$, shifting downward with increasing voltage. Note that $T_c(V)$ reflects the voltage-controlled transition onset rather than the intrinsic bulk $VO_2$ transition temperature ($T_{\mathrm{IMT}} \approx 340$ K, SM Sec. S1). Numerical simulations of the Joule-coupled resistor-network model reproduce this crossover very well (Fig. 2(b)). In Fig. 1(c-f) we illustrate how this transition develops, resulting from the combination of the percolative nature of the IMT together with the effect of Joule heating from metallic domains, see the figure caption for a frame-by-frame explanation. One can quantitatively capture the change from a continuous R–T to a transition with an abrupt jump via a mean-field model of Joule heating, see SM Sec. S3, indicating that the abrupt transition temperature is governed by overall sample heating. However, this model is not expected to capture the R–T curve near the transition, nor the time-dependent changes, both of which are governed by the dynamics of local IMT abrupt switching events.

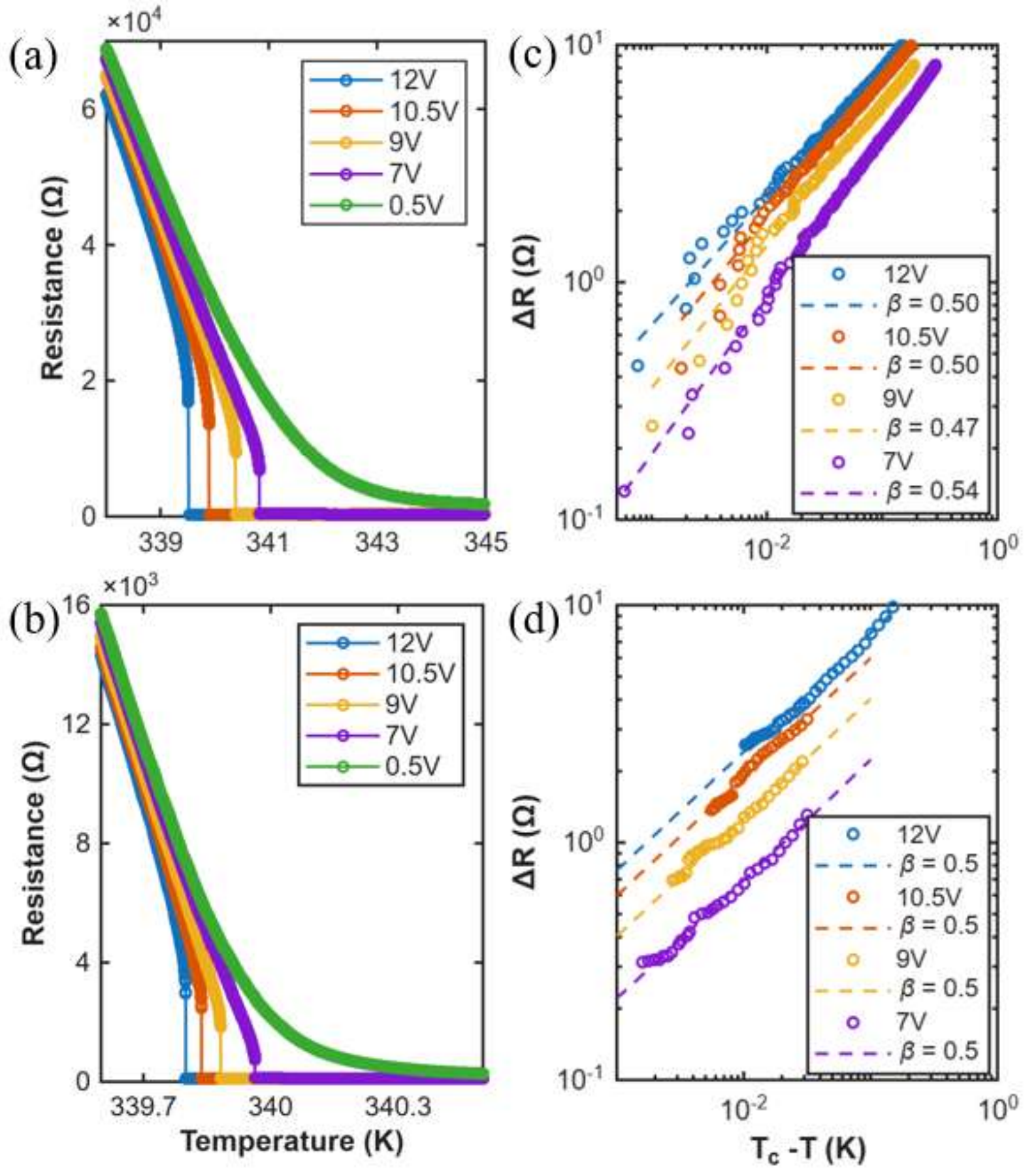

FIG. 2. R-T results of (a) the $VO_2$ network device and (b) simulation model, under different bias voltages. The experimental results were taken at a temperature ramp rate of 0.1 K/min. A change from a smooth Type II transition (green) into an abrupt mixed order transition is observed at around 5.5 V, as well as the decrease of the transition Tc. (c)-(d) power law behavior near the transition of (c) network device and (d) simulation model, for different bias voltages, with exponents approaching $\beta \sim 0.5$.

To analyze the critical behavior of these abrupt (Type I) transitions, we use the scaling form expected near a mixed-order transition, which predicts power-law behavior near criticality [31,40]:

$$\Delta R \propto (\Delta T)^{\beta},$$

where $T_C$ is the temperature of the abrupt transition, $\Delta R = R(T) - R_C$ is the difference between the resistance before and at criticality ($R_C = R(T_C)$), $\Delta T = |T - T_C|$ is the reduced temperature and $\beta$ is the critical exponent. Scaling parameters are extracted from the data points immediately preceding the abrupt drop. Because discrete sampling and experimental precision prevent direct measurement of the precise onset of $T_C$, $T_C$ is determined by bounding values across the drop interval. A similar extraction procedure is applied to the discretized simulation outputs, where $T_c$ can be constrained with higher precision. As shown in the log-log plots in Figs. 2(c) and 2(d), both experimental and numerical data yield a critical exponent $\beta \approx 0.5$, consistent with the mean-field universality class expected for interdependent networks [42,43] .

We applied a measurement protocol aimed at assessing the cascading nature expected in interdependent networks. Its hallmark is a long resistance plateau at constant temperature, which, depending on whether the network is below or above the critical point (set by the temperature and the applied voltage), either stabilizes or ends in a network collapse. The network was cooled to 290 K and heated at 1 K/min under an applied voltage of 0.5 V until reaching a specific preset temperature (and specific resistance). After a 750 s thermal stabilization, the applied voltage was stepped to a higher value (e.g., $V = 10$ V as shown in Fig. 3(a); see SM Sec. S4 for other voltages), and the system evolved isothermally over an extended duration, up to 20 hours, to monitor self-heating-induced cascading transition dynamics. See SM Sec. S2 for additional details of the measurement.

Figure 3(a) illustrates the resulting temporal resistance evolution for an applied voltage of 10 V and different set temperatures (see SM Sec. S4 for additional voltages). Near criticality ($T \approx T_c$), the latency prior to the abrupt resistance drop, defined as the plateau time $\tau_p$, grows sharply. At and below criticality ($T \leq T_c$), $\tau_p$ exceeds the measurement window and is treated as infinite. Rather than a simple exponential decay, in the vicinity of the critical temperature $\tau_p$ follows power-law scaling with reduced temperature, $\tau_p \propto |T - T_c|^{-\zeta}$ with $\zeta \approx 0.5$, as demonstrated in Fig. 3(c), and in agreement with previous reports of $\zeta$ in cascading networks [28,34,40]. The same exponent characterizes delayed switching in driven systems [44]; here, however, the plateau exceeds any thermal time constant of the device by orders of magnitude and consists of discrete switching steps (SM Sec. S5), pointing to a random cascade rather than a deterministic slowing-down. This indicates that the critical dynamics of the transition govern not only the stationary resistance value but also the temporal delay of the transition onset. As expected, the data depart from the power law as the system moves away from criticality.

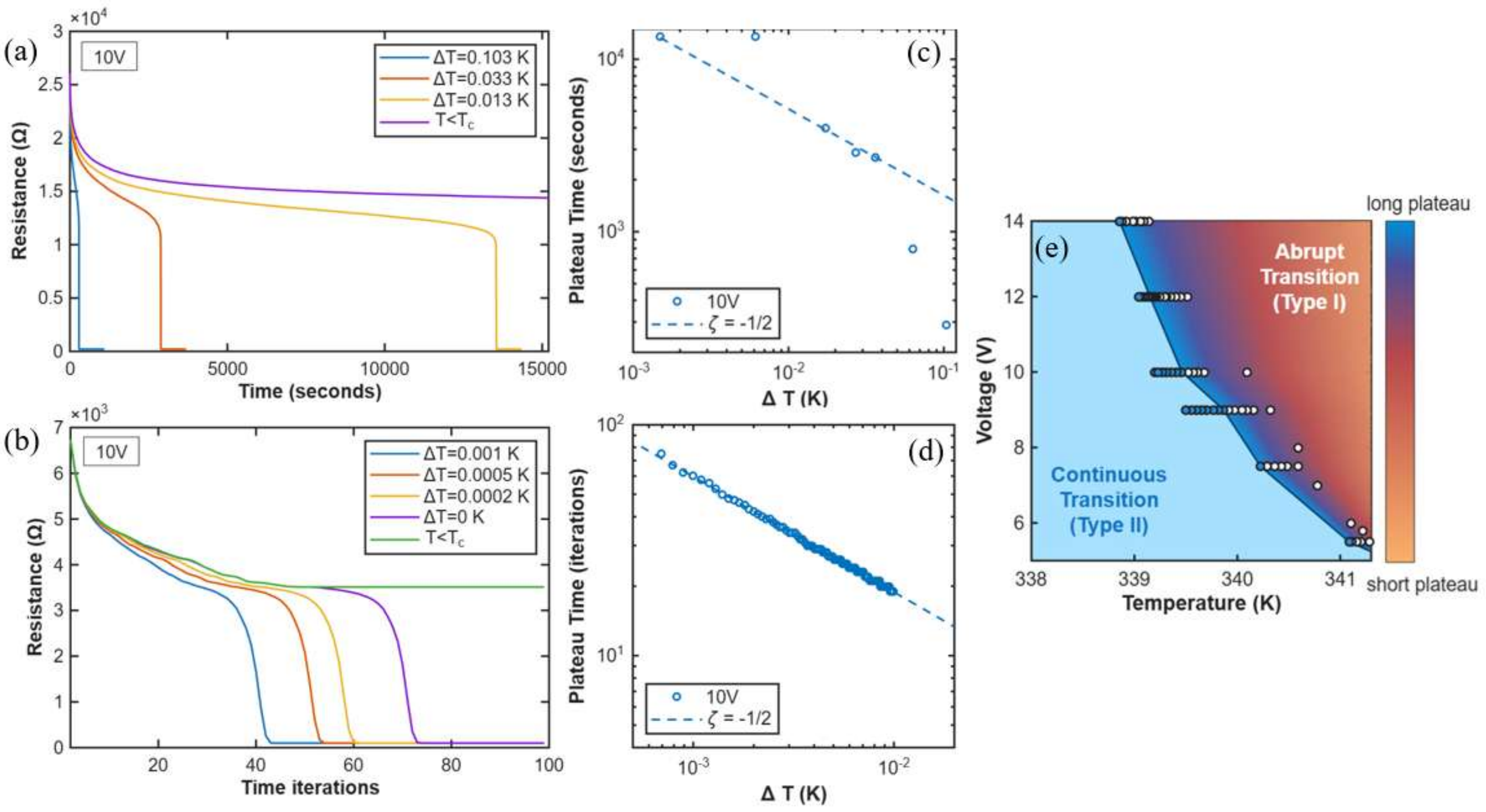

FIG. 3. (a) Experimental resistance $R$ versus time for various temperatures near $T_c = 339.937K$ at an applied voltage of 10 V. Near criticality ($T \approx T_c$), the delay prior to the resistance-drop (plateau time) increases, exceeding the measurement duration at $T_c$. (b) Simulated R vs. time-iterations for a network model showing analogous behavior; here, the plateau diverges only above $T_c$ due to exact determination of $T_c$. (c) Experimental power-law scaling of the plateau time versus reduced temperature ($T - T_c$), with exponent $\zeta \sim 0.5$. (d) Simulated power-law scaling of the iteration plateau time versus reduced temperature, yielding a critical exponent $\zeta \sim 0.5$. (e) Phase diagram of transition types versus voltage and temperature. At low voltages and temperatures, the transition is continuous (Type II). Above a temperature-dependent threshold voltage, it becomes abrupt (Type I), with plateau times shortening away from criticality.

"In the simulations [Fig. 3(b)], collapse occurs only for T > Tc, and the plateau length grows without bound as T approaches $T_C$ from above; the sharper divergence reflects the exact $T_C$ accessible only in simulation, since it eliminates thermal uncertainties present in the experimental setup. Figure 3(d) confirms the power-law scaling of the simulated iteration plateau time, yielding a critical exponent consistent with experiment.

By mapping $\tau_p$ across a range of applied voltages and ambient temperatures, we construct the dynamic phase diagram shown in Fig. 3(e). At low voltages and temperatures, the network undergoes a continuous (Type II) transition. Above a temperature-dependent critical threshold $V_c(T)$, the transition becomes abrupt (Type I), where the plateau time diverges along the critical boundary line $V_c(T)$ and monotonically decreases far from criticality. Thus, the phase map delineates the boundary between the continuous (Type II) and abrupt (Type I) regimes.

Using canonical branching process theory [45], we analyzed the transition by calculating its branching factor $(BF)$; we refer to metallic segments in the network as infected and define the branching factor as the average number of segments infected (in the next time step) by a single infected segment. This analysis was performed both for the experimental results and for the simulations. For a branching process, such as an epidemic spreading, the critical threshold is 1. That is, for $BF < 1$, the process dies out, as there are not enough newly infected segments to sustain it. For $BF > 1$, a cascading process occurs, where more and more segments are infected and turn metallic, infecting even more segments themselves in an exponential manner, pushing the network transition. Therefore, BF = 1 is the critical point at which each infected segment infects, on average, exactly one other segment, so the infection neither dies out nor grows exponentially and the spreading time diverges (in a large enough system). This can be seen in Fig. 4, where the branching factor is plotted against time for both above and below criticality, for the experimental and simulated results.

It can be clearly seen that the branching factor exceeds 1 (beyond criticality) much earlier than the actual transition of the network. In all collapsing runs, once the branching factor exceeds 1 it does not return below unity, and the network collapse follows, up to thousands of seconds later. However, if the branching factor is below 1 for the entire measurement, no transition occurs within the measurement window. Thus, the branching factor can be seen as a predictor, capable of identifying criticality in the system long before the actual transition.

Using the experimental data, we could not apply the canonical branching-process estimator of the branching factor [30,46,47], which requires counting discrete events and assigning them to successive generations. Switching events are resolvable in the resistance trace, but we cannot tell which event triggered which, so no ancestor-to-descendant ratio can be formed. Nor is the cascade separated into distinct avalanches that would define a natural time bin. Moreover, the activity decays continuously over three decades. Compounding this, the number of insulating domains that turned metallic within a time window is not directly measurable, since the conversion between a resistance change and a switched area is set by the effective-medium response of the film and is itself a function of how much has already switched [5,48,49]. We therefore measure instead the rate at which switching occurs and ask how that rate evolves. We define the switching activity $a(t) = dG/dt$, extracted from the conductance $G = 1/R$ in logarithmically spaced time bins, $G$ being the quantity extensive in the number of switched regions. The local activity exponent $\gamma(t) = -d(ln\, a)/d(ln\, t)$ was obtained from a smoothing spline of $ln\,(a)$ versus $ln\,(t)$. We define $\eta(t) = 2^{-\gamma(t)}$, the factor by which the switching activity changes per doubling of elapsed time, as an effective branching factor. Here $\eta = 1$ is the point at which the activity becomes scale-free in time, which corresponds exactly to BF = 1 of the underlying branching process, for any generation time (SM Sec. S5); away from $\eta = 1$, $\eta$ is an observable of the activity record rather than a generation-by-generation branching ratio. Thus, $\eta$ exceeding 1 signifies the onset of the runaway. A full explanation is given in SM Sec. S5.

Figure 4 plots the temporal evolution of $\eta(t)$ together with R(t) for experimental measurements below criticality, where BF < 1 at all times [Fig. 4(a)]; close to criticality, where BF is near 1, resulting in an extended plateau before an avalanche [Fig. 4(b)]; and a case where the temperature is higher than $T_C$, resulting in a short delay before the system collapses [Fig. 4(c)]. In Fig. 4(d) we present the numerical analysis of the BF for the three different cases. Here, since there is no experimental noise, and switching events can be directly assessed, there is no issue with extracting the branching factors. The crossing of BF = 1 corresponds, as in the experiment, with the system's critical behavior. The early threshold crossing of BF>1 is always followed by the runaway transition, whereas trajectories where $\eta(t)$ and BF are smaller than 1 (subcritical) throughout the measurement remain trapped in the insulating phase. Tracking the branching factor thus provides an early predictor of the network heading towards a global resistance collapse.

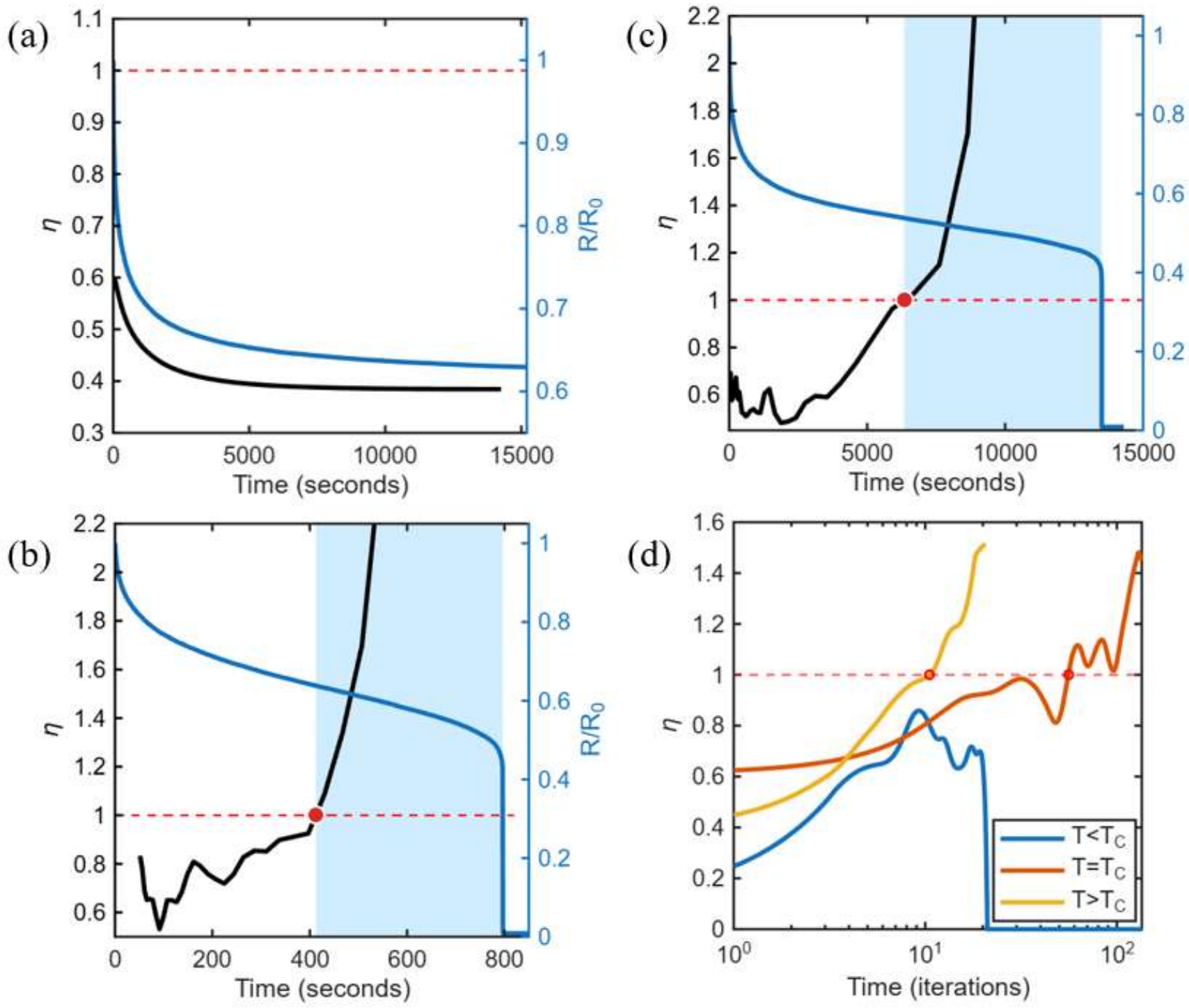

FIG. 4. (a-c) Experimental branching factor versus time ($\eta$, black lines and left axis) shown alongside the normalized resistance plot (R/R(t=0), blue line and right axis). Critical threshold of 1 is indicated by the dashed red horizontal line. The red circles highlight the time at which the branching factor crosses 1 (if present), and the light-blue area marks the interval between this crossing and the resistance drop. (a) Below the critical temperature ($T < T_c$), no collapse. (b) $T > T_C$ and close to it. Long plateau leading to collapse. $\eta$ crosses 1 about 7150 s before the collapse. (c) $T \gg T_C$ showing $\eta$ crossing 1 and a shorter plateau. (d) Branching factor from the numerical simulations for three different temperatures, collapse (if present) is marked, time steps are on a logarithmic scale: Blue line - $T < T_c$ and BF does not cross 1. Red line - $T$ above $T_C$ showing long plateau till BF crosses 1. Green line - $T \gg T_C$ showing BF crossing 1 and network collapse in a shorter time.

## III. DISCUSSION

The close agreement between the experiments and the computational simulations, as evident in the abrupt IMT in Fig. 2(a,b), the incubation plateaus in Fig. 3(a,b) and the branching factors in Fig. 4(a-d) indicates that the Joule-coupled network model captures the underlying mechanism by which the voltage driven IMT in $VO_2$, and possibly field-driven IMTs in general, develop. Examination of these systems near the critical temperature reveals that the abrupt, mixed-order collapse of the resistance (a macroscopic change) is not an instantaneous event, but the culmination of a long chain of individually negligible microscopic switching events that, given sufficient time, propagate until the entire macroscopic state has changed. Moreover, the fact that our minimal numerical model reproduces the main experimental findings indicates that the transition is governed by the overall thermodynamic properties of the system. That is, the dynamics follow cascading-network theory and do not require extensive local heat transport models in order to understand the systems' behavior.

The cascading nature of the transition is evident from the temporal behavior as we modify, even slightly, the temperature of the system (for a given voltage). For example, in Fig. 3(a), changing the dwell temperature by 20 mK increased the plateau from ~3000 seconds to over 13,000 seconds. There are two important outcomes. First, there is a clear critical temperature below which there is no network collapse, no matter how long the system dwells. Second, when the system is above $T_C$ the closer the temperature is to $T_C$, the more prolonged the transition becomes before the system collapses. That is, rather than collapsing at once, the resistance changes slowly in a long metastable plateau of minor changes that only after a considerable timespan leads to the eventual collapse, as seen in Fig. 3. While previous studies either ignored the 'delayed' phase transition or reported it on time scales between nanoseconds and milliseconds, our study shows that when approaching criticality this transition can take well over ten thousand seconds. Moreover, it shares critical exponents with interdependent-network models, suggesting that what we report for $VO_2$ should apply broadly to physical systems in which dissipation provides positive feedback [31,33,34,39].

Following the standard definitions used in the field of cascading and interacting networks [24,25,35,50], the second layer of interaction (the 'dependency') in $VO_2$ is controlled by the magnitude of the applied voltage and mediated by Joule heating; it drives a potentially long cascade and a mixed-order phase transition. That is, the transition is neither a completely abrupt first order transition nor is it a continuous second order transition. Rather, it begins with a curve akin to a second order transition (with $\beta = 0.5$ as shown in Fig. 2 (c-d)) and culminates in a first-order jump.

The considerably long time it takes the system to conclude the transition and reach its new steady state indicates that the process is indeed a long chain of microscopic changes in which individual links are affected and then affect other individual links until the entire system changes. This is best seen by our analysis of the branching factor that approaches 1 the closer we are to the critical temperature, with a branching factor >1 above $T_C$ and <1 below $T_C$ (where no collapse occurs). We were also able to determine, well before the actual collapse, that the system will indeed make the transition, giving us a "predictor" tool to know that such a collapse will happen sometime in the future, simply by observing that the branching factor crosses the threshold of 1. Put simply, beyond being a diagnostic tool of the transition criticality upon approaching the critical temperature, the branching factor analysis is a predictor of the system's final state —whether it will undergo a catastrophic collapse or a gentle relaxation. When the branching factor is below unity, the system activity is subcritical, i.e., each domain affects less than one domain and the cascading is expected to die out, so the plateau remains stable.

Furthermore, the fact that both the experiment and the simulations show the same power laws near criticality as shown in Figures 2(c,d) and 3(c,d) is a strong indication that the IMT of the $VO_2$ network is part of a large class of systems, all governed by similar principles and consistent with studies in network science [33,42,43].

Ongoing work will shed more light on the influence of the voltage, on the process of cooling compared to that of heating, on whether the size effect also follows the same trajectory as that of cascading networks, and on effects from different system geometries.

In summary, we have shown that the voltage-driven insulator–metal transition of a patterned $VO_2$ lattice is a mixed-order cascading transition of a single network. Above a threshold bias, the Joule power, which grows as links turn metallic, acts as a dependency coupling within the resistor network and converts the continuous, percolative thermal transition into an abrupt collapse. On approaching the critical temperature, the collapse is preceded by a metastable plateau whose lifetime diverges as $\tau_p \propto |T - T_c|^{-\frac{1}{2}}$ and exceeds $10^4$ s, orders of magnitude longer than any thermal time scale of the device. The resistance approaches the jump with an exponent close to $\beta = 1/2$, and an effective branching factor of the switching activity crosses unity well before the collapse, providing a predictor for the collapse. A minimal Joule-coupled resistor-network model reproduces these signatures, which coincide with those of interdependent networks. Genuine interdependence between two distinct networks is therefore not required: dissipative self-coupling within one network suffices.

We conclude that the results presented in this study strongly indicate that what was previously studied and analyzed mainly theoretically and numerically in the field of network science (which normally involves large scale systems such as infrastructure, climate networks, economy and more) applies to much smaller, physical and controlled systems as well. This bridge between the two fields opens possibilities in both research and application, by using the insights learned in complex networks to understand the dynamics of the physical ones. Conversely, it provides a controlled environment in which to test network theories that were previously difficult to test experimentally.

## Acknowledgements

A.S. acknowledges the support of the Israel Science Foundation No. 1499/23 and MOST No. 6068. S.H. acknowledges the support of the Israel Science Foundation (Grant No. 201/25), the EU H2020 DIT4TRAM, EU H2020 project OMINO (Grant No. 101086321), and the VATAT National Foundation for Climate and Power Grid networks and the Israel Ministry of Innovation, Science and Technology (grant number 01017980), for financial support.

## Conflict of Interest

The authors declare no conflict of interest.

## Data Availability Statement

The data that support the findings of this study are available from the corresponding author upon reasonable request.

# Supplementary Material

## Interdependent-Network Criticality without a Second Network: Cascading Collapse of the Joule-Coupled Insulator–Metal Transition in $VO_2$

Ouriel Gotesdyner[1], Michal Wasserman[1,2], Hodaya Kaster[1,2], Yosi Abulafia[1,2], Avital Fried[1,2], Bnaya Gross[3], Shlomo Havlin[1], and Amos Sharoni[1,2,†]

[1]Department of Physics, Bar Ilan University, Ramat-Gan, 5290002, Israel.

[2]Institute of Nanotechnology & Advanced Materials, Bar Ilan University, Ramat-Gan, 5290002, Israel.

[3]Department of Mathematics, Bar Ilan University, Ramat-Gan, 5290002, Israel.

[†]Corresponding author.

## S1. Device preparation and measurement

The high-quality 65-nm-thick $VO_2$ thin films were grown on (012) $r$-cut sapphire ($\alpha$-$Al_2O_3$) substrates using reactive radio-frequency (rf) magnetron sputtering from a 2-inch $V_2O_3$ target (99.9% purity, ACI Alloys, Inc.). Prior to deposition, the high-vacuum chamber was evacuated to a base pressure below $1.2 \times 10^{-8}$ Torr. The substrates were thermally outgassed at 700°C and subsequently ramped down to the growth temperature of 550°C at a rate of 5 K/min. Sputtering was performed at rf power of 140 W in an $Ar/O_2$ gas mixture consisting of 26 sccm Ar and 0.4 sccm $O_2$, maintaining a total operating pressure of 3.5 mTorr. Following deposition, the films were cooled to room temperature under the same $Ar/O_2$ mixture to preserve oxygen stoichiometry and prevent the formation of oxygen vacancies.

The two-dimensional resistor networks were patterned using high-resolution electron-beam lithography (EBL) on a PMMA resist layer and developed in a methyl isobutyl ketone and isopropanol mixture (MIBK:IPA, 1:3). The patterned geometry, comprising wire widths of 500 nm, a pitch of 1 $\mu$m, and a total active area of $200 \times 200\ \mu m^2$, was transferred to the $VO_2$ layer via reactive ion etching (RIE) utilizing a $SF_6$ plasma at a flow rate of 50 sccm, a chamber pressure of 5 mTorr, and plasma power of 100 W. Residual resist was stripped by Acetone.

In a subsequent alignment step, electrical contact pads were defined via EBL using a bi-layer resist stack for clean lift-off. Metallic leads consisting of 10 nm Cr / 80 nm Au were deposited by electron-beam evaporation, followed by lift-off.

To establish the intrinsic transport characteristics of the pristine $VO_2$ layer prior to nano-patterning, temperature-dependent resistance measurements, $R(T)$, were performed on an unpatterned region of the same chip. Measurements were carried out during continuous heating and cooling cycles across the insulator-metal transition (IMT) at a controlled sweep rate of 1 K/min.

Figure S1 shows the characteristic thermal hysteresis loop of the resistance across the IMT. The unpatterned film exhibits an abrupt resistance change spanning 4 orders of magnitude between the low-temperature insulating phase and the high-temperature metallic phase. The transition temperature is 339 K and the thermal hysteresis width is $\approx 10K$. The large resistance change confirms the high quality of the film on the $r$-cut sapphire substrate. This baseline behavior serves as a benchmark confirming that the discrete switching events and percolation dynamics observed in the 2D resistor networks originate from spatial confinement and network connectivity rather than film quality.

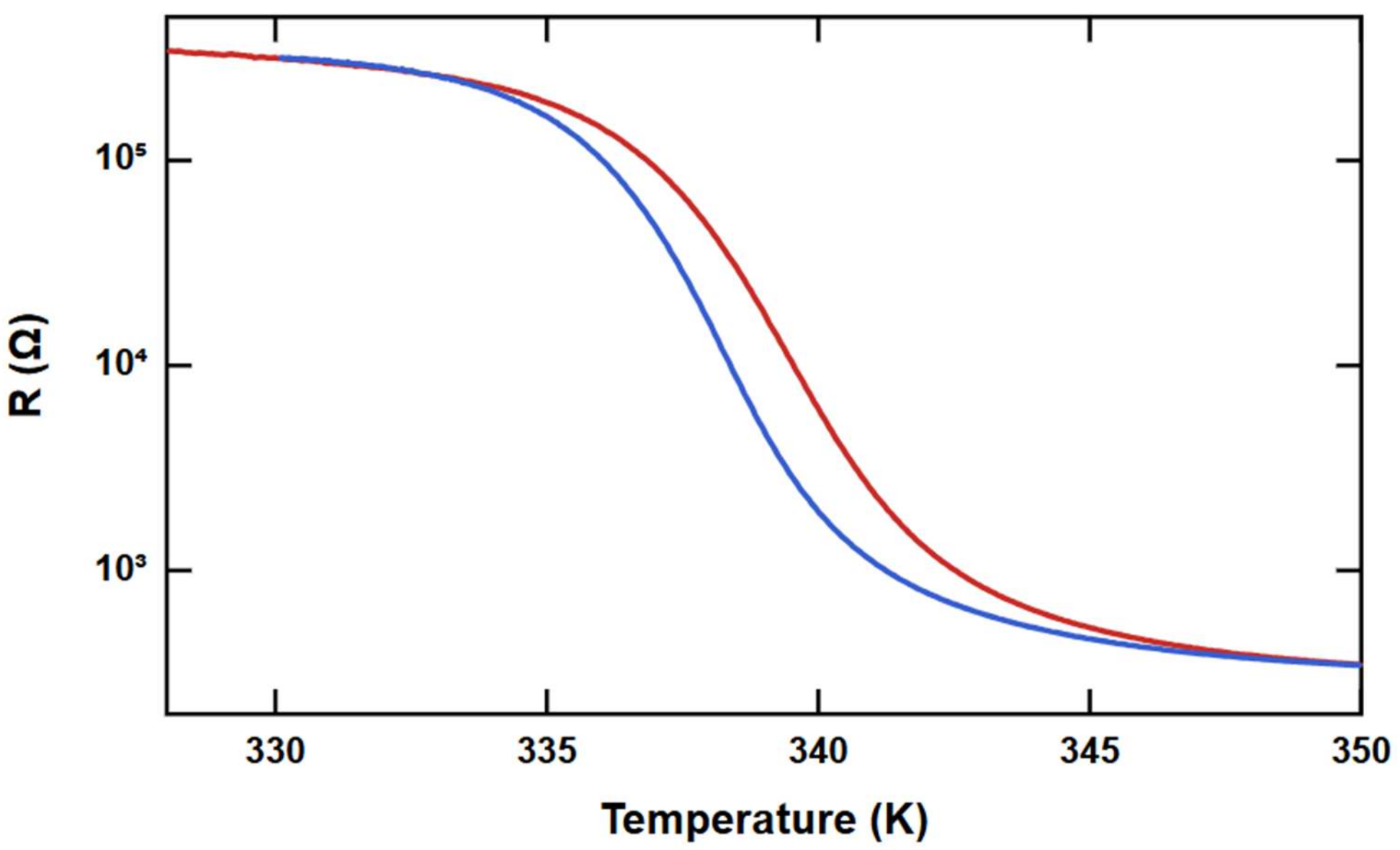

FIG. S1 Thermal hysteresis of unpatterned $VO_2$ thin film. Temperature-dependent resistance $R(T)$ recorded during heating (red) and cooling (blue) cycles across the metal–insulator transition.

## S2. Measurement system

S2.1 Measurement system for control of the bath temperature

Measurements of the $VO_2$ network require reproducible temperature control of the sample, while allowing the sample temperature to differ from the thermal bath temperature. We therefore control the temperature of the thermal bath while keeping a moderate thermal coupling between the bath and the sample. This coupling allows the bath to set the sample temperature, while letting the network self-heat without changing the bath temperature; if the sensor were on the sample, the controller would respond to self-heating by increasing the cooling, i.e., it would act as a negative feedback on local heating. Whereas the temperature sensor is usually placed on the sample, here we deliberately place it away from the sample, on the bath.

In addition, since we are running experiments for many hours under a constant temperature, high global temperature stability is required, to hold the system precisely within the narrow phase-coexistence regime and eliminate baseline drift.

To satisfy both requirements, the sample stage was configured as pictured in Fig. S2-1. The base consists of a high-purity bulk copper heat sink featuring a central raised pedestal, surrounded by a stationary PCB breakout board with pin-socket receptacles. The sample chip is mounted onto an exchangeable PCB chip carrier with an exposed copper backing, mechanically and thermally coupled using a thin layer of thermal GE-varnish, and electrically connected via wire bonds to perimeter pads. The chip carrier plugs into the socket board, seating its copper underside against the pedestal via a layer of high quality thermal grease.

Crucially, the control thermometer (Pt-1000 sensor) and Peltier-based heater/cooler are mounted directly to the bulk copper base rather than adjacent to the sample die. Combining the remote sensor placement with the large heat capacity of the copper block establishes a stable thermal bath while preventing the controller from actively compensating for rapid, localized percolation events. The entire assembly is enclosed in a chamber with a number of insulating layers, to prevent changes in the room temperature from affecting the setup.

To evaluate the thermal stability of the measurement stage, benchmark characterization runs were performed across multiple setpoints near $T_c$ (Fig. S2-2). Following the initial heating ramp from room temperature (Fig. S2-2(a)), the heat-bath temperature approaches the target setpoint smoothly and settles within approximately 100 s without noticeable overshoot (Fig. S2-2(b)).

Under steady-state conditions (Fig. S2-2(c)), the stage maintains excellent long-term stability over observation windows exceeding $1.5 \times 10^4$ s ($>$ 4 hours). The thermal noise measured at the copper base has an RMS deviation of 0.76 mK and a peak-to-peak variation below 5.56 mK, confirming that global temperature fluctuations are effectively eliminated as a source of experimental error during prolonged percolation measurements.

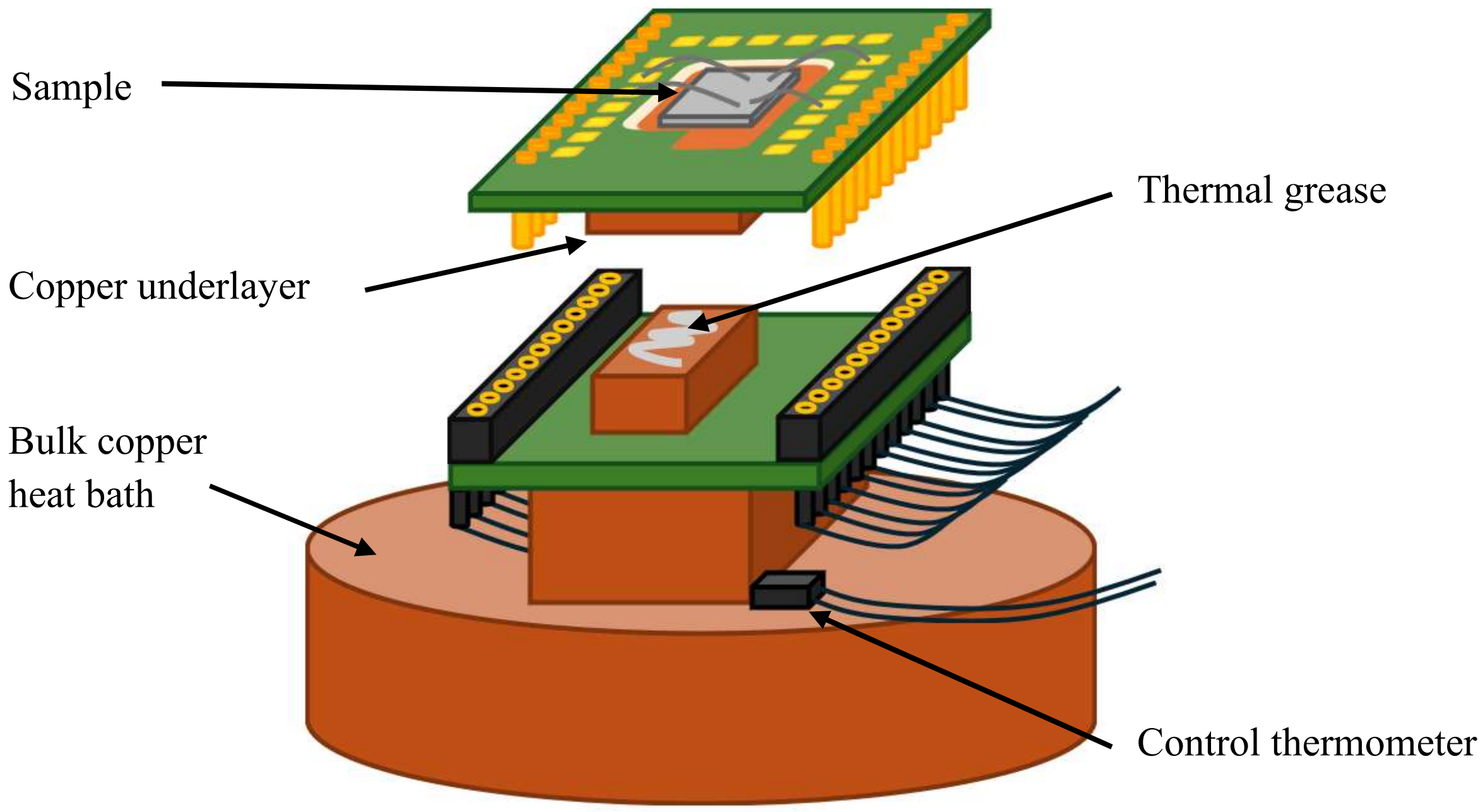

FIG. S2-1. Schematic of the measurement stage assembly. To maintain precise global temperature stability without clamping local Joule-heating dynamics, the control sensor is anchored to the massive copper heat sink base rather than adjacent to the sample, thermally decoupling fast local percolation events from the controller feedback loop.

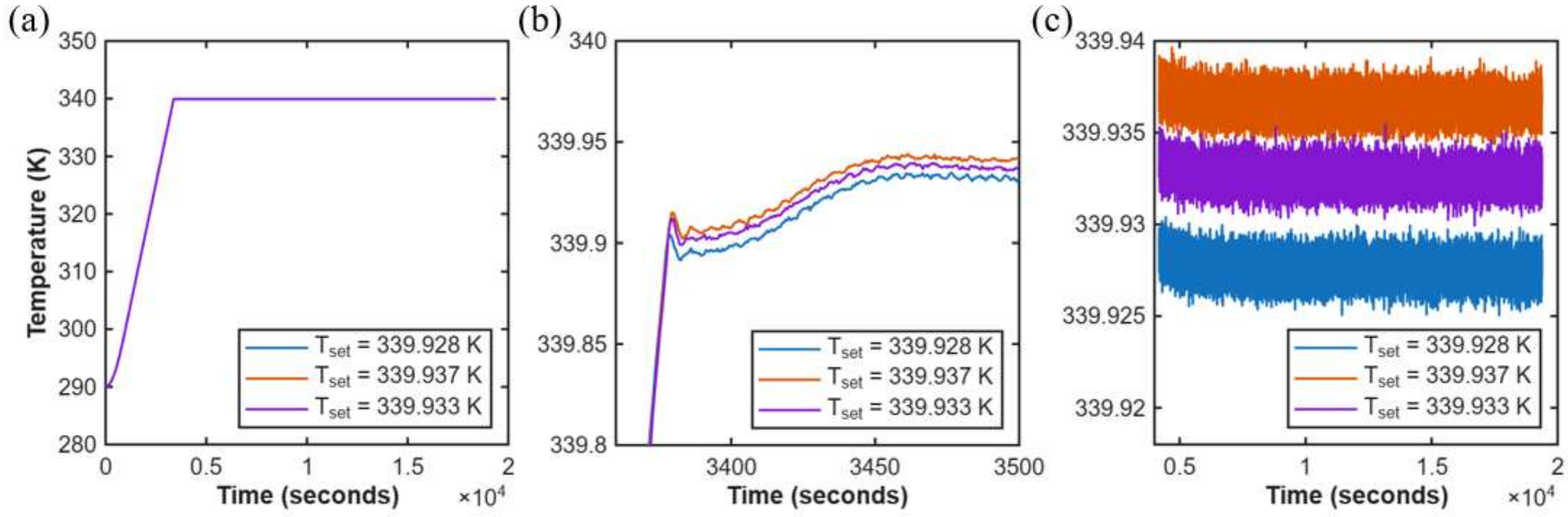

FIG. S2-2. Thermal stability characterization of the copper heat bath. **(a)** Full temperature trajectory from room temperature to the target setpoint near $T_c$. **(b)** Detailed transient behavior during setpoint acquisition, illustrating smooth settling and minimal overshoot. **(c)** Steady-state thermal stability recorded over $> 1.5 \times 10^4$ s for three representative setpoints separated by millikelvin increments, exhibiting sub-millikelvin RMS noise ($< 0.76$ mK).

### S2.2 Dependence on ramp rate

The dependence of the insulator–metal transition (IMT) and the resulting resistance-temperature ($R$-$T$) hysteresis on the temperature sweep rate was measured in the $VO_2$ networks under a constant bias voltage. Representative curves acquired under 12 V and at sweep rates ranging from 0.1 to 2 K/min are shown in Fig. S2-3.

Across this parameter space, variation in the sweep rate produces only marginal changes in the overall switching characteristics. While both the quasi-static sweep rate of 0.1 K/min and the faster rate of 2 K/min, shift the transition toward slightly higher temperatures compared to rates of 0.5 and 1 K/min, the overall behavior is the same. We find that the small shifts of the transition temperature correlate with minor changes in the low-temperature resistance, and can be minimized by performing several heating–cooling cycles prior to the measurements, as was done in all experiments.

A sweep rate of 1 K/min was used for the ramps preceding the isothermal dwells (Fig. 3 and Sec. S4), and 0.1 K/min for the R–T curves of Figs. 1(a) and 2.

## S3. Mean-field model of the voltage-dependent transition

We used a mean field model to fit R vs. T curves at different voltages, assuming only Joule heating of the effective sample resistance, solved self-consistently. All bias voltages are described by one mean-field model with a single parameter set. The film is treated as a two-phase random medium whose local transition temperatures are normally distributed about a transition temperature, $T_{IMT}$, with standard deviation $\Delta T$, so the metallic area fraction at film temperature $T_e$ is

$$f(Te) = ½ [1 + erf((T_e - T_{IMT}) / \sqrt{2}\ \Delta T)] \qquad \text{(S3.1)}$$

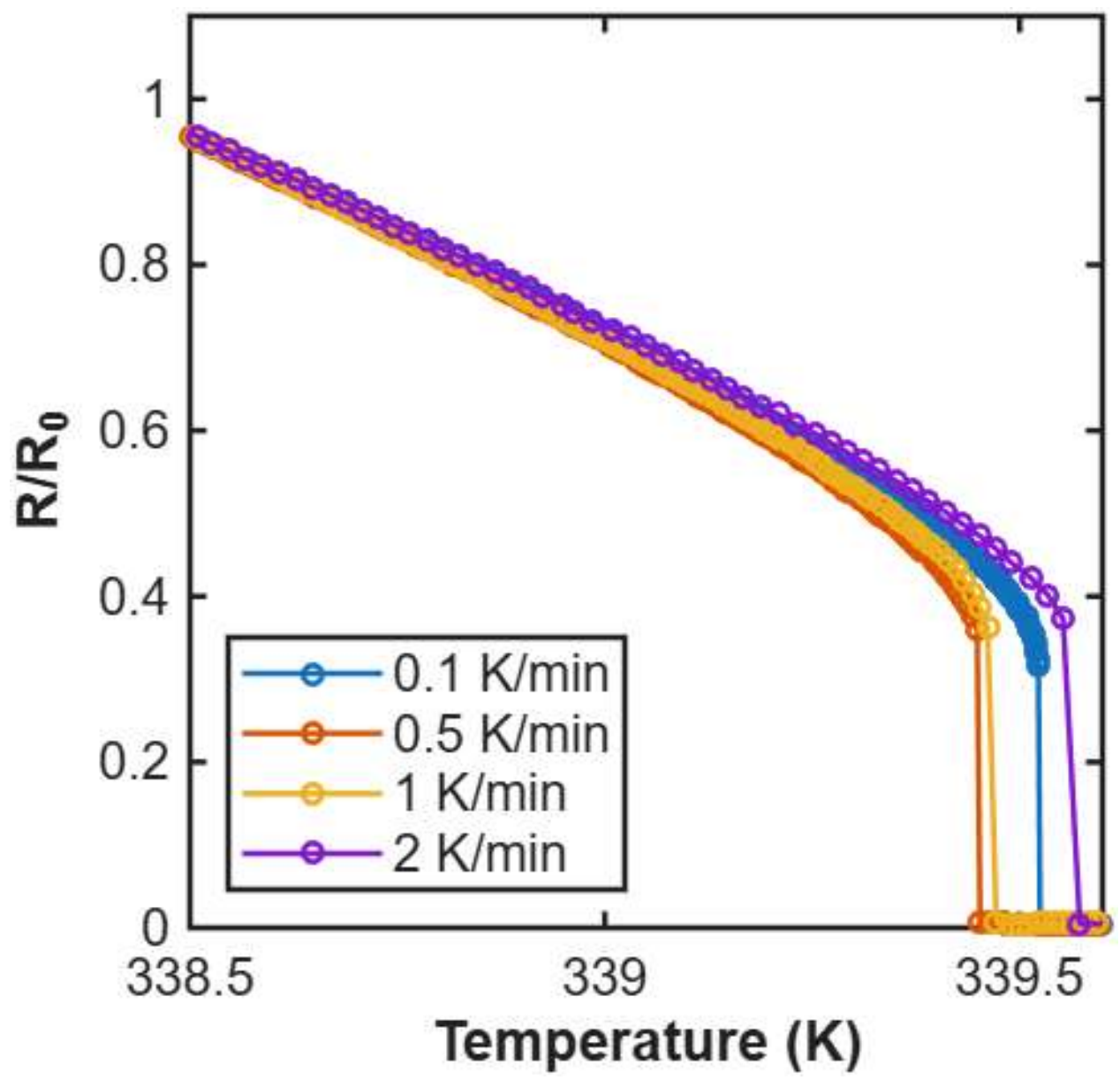

FIG. S2-3: Resistance measurements for different temperature ramp rates. Temperature-dependent resistance $R(T)$ of a $VO_2$ network measured under an applied bias of 12 V at ramping rates of 0.1, 0.5, 1.0, and 2.0 K/min.

The insulating regions are thermally activated, $R_I(Te) = R_0\ exp(E_0/T_e)$, the metallic ones have a temperature-independent $R_M$, and the resistance is assessed via a two-dimensional effective-medium expression [1,2],

$$Reff = \frac{1}{2}\left[-b + \sqrt{b^2 + 4R_I R_M}\right], \quad b = (2f - 1)(R_I - R_M) \quad \text{(S3.2)}$$

which gives $R_{eff} = R_I$ at $f = 0$, $R_M$ at $f = 1$.

The bias enters in one place only, through Joule self-heating: dissipation raises the film above the stage temperature by an amount set by a single thermal coefficient $a_0$, $T_e = T + a_0 V^2/R$. Since $R$ itself depends on $T_e$, this closes into a self-consistency condition solved at each temperature and bias,

$$R = R_{eff}(T + a_0 V^2/R). \qquad \text{(S3.3)}$$

The fitting protocol is done as follows.

**(i) Find parameters from the 0.5 V curve.** At 0.5 V the dissipation is negligible ($a_0 V^2/R < 0.05$ K over the whole transition), so Eq. (S3.3) reduces to $R = R_{eff}(T)$ and the measured curve probes the material response alone. The five parameters $T_{\mathrm{IMT}}$, $\Delta T$, $E_0$, $R_0$, $R_{\mathrm{M}}$ (the metallic fraction versus temperature plus the two branch resistances) come from fitting log($R$), so the beginning and end of the transition have similar weights, resulting in: $T_{\mathrm{IMT}}$ = 341.99 K, $\Delta T$ = 7.04 K, $E_0$ = 2371 K, $R_0$ = 136 Ω and $R_{\mathrm{M}}$ = 298 Ω, reproducing the 0.5 V curve very well, see blue line in Fig. S3.

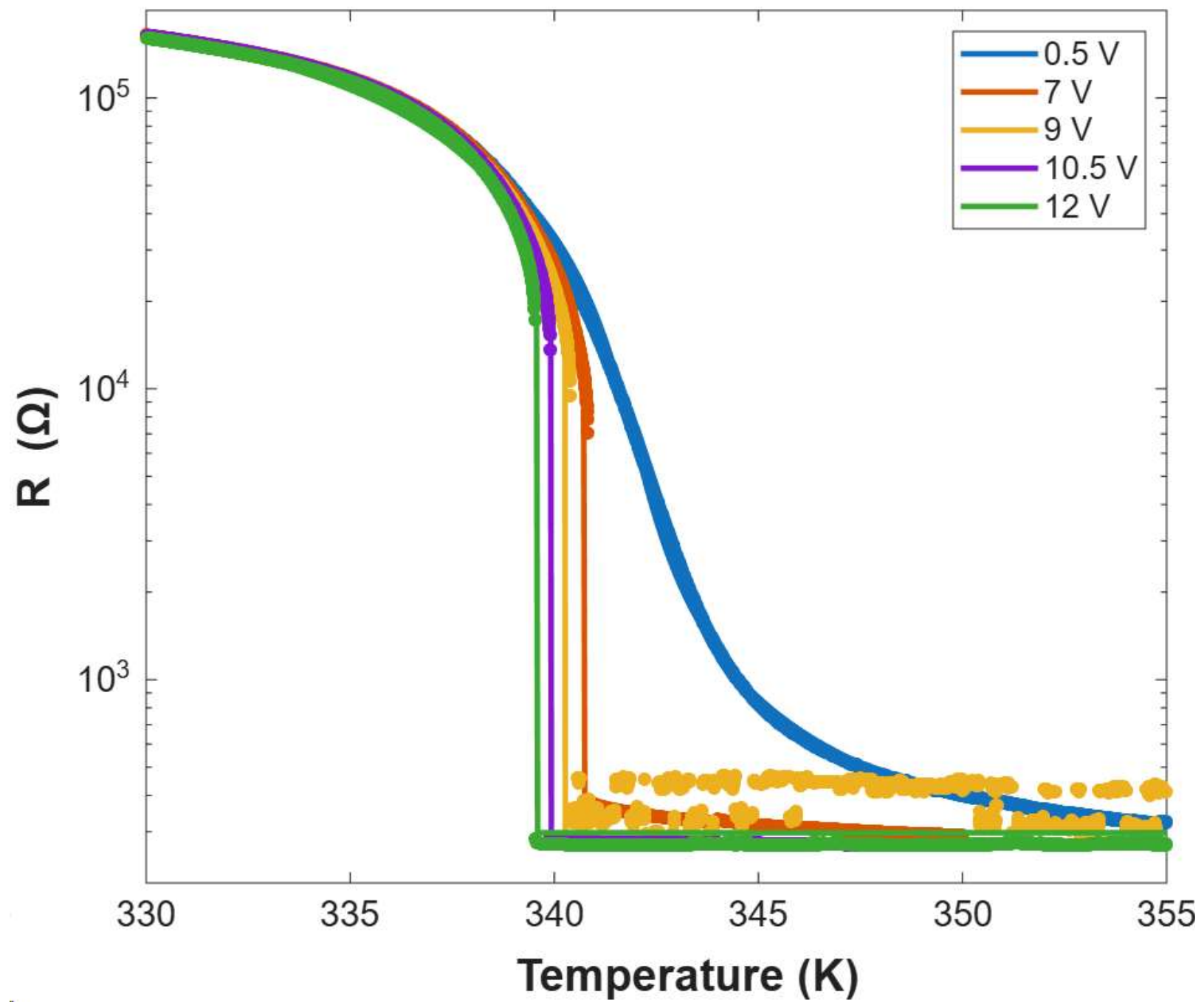

FIG. S3: Comparing data to mean-field model. R vs. T measured for different voltages are plotted as circles, voltages appear in legend. The mean-field model is plotted as a continuous line for each voltage. See text for details.

**(ii) Finding thermal coefficient from one biased curve.** With those five constants held, the single remaining parameter $a_0$ is fitted to the 7 V curve, giving $a_0 = 180\ K\ \Omega/V^2$.

**(iii) Calculating other voltages with no free parameters.** Every remaining curve in Fig. 2 is then a prediction. Eq. (S3.3) is solved with the six parameters already fixed and nothing is adjusted per voltage.

The mean-field model reproduces well the temperature at which the resistance drops for the three remaining measurements. These results indicate that it is reasonable to assume that the heating of the sample is global, as we do in the numerical simulations. We note in passing that we analyzed the $\beta$ parameter (slope of $\log{(R - R_c)}$ vs $\log{(T_{IMT} - T)}$, not shown here) for the mean-field results, and find $\beta \sim 0.57 - 0.6$ as the best fit to a range similar to the experimental data. $\beta$ approaches 0.5 only if we look at the curve up to 5 mK from the critical temperature, while the experimental results show the fit extends to above 100 mK.

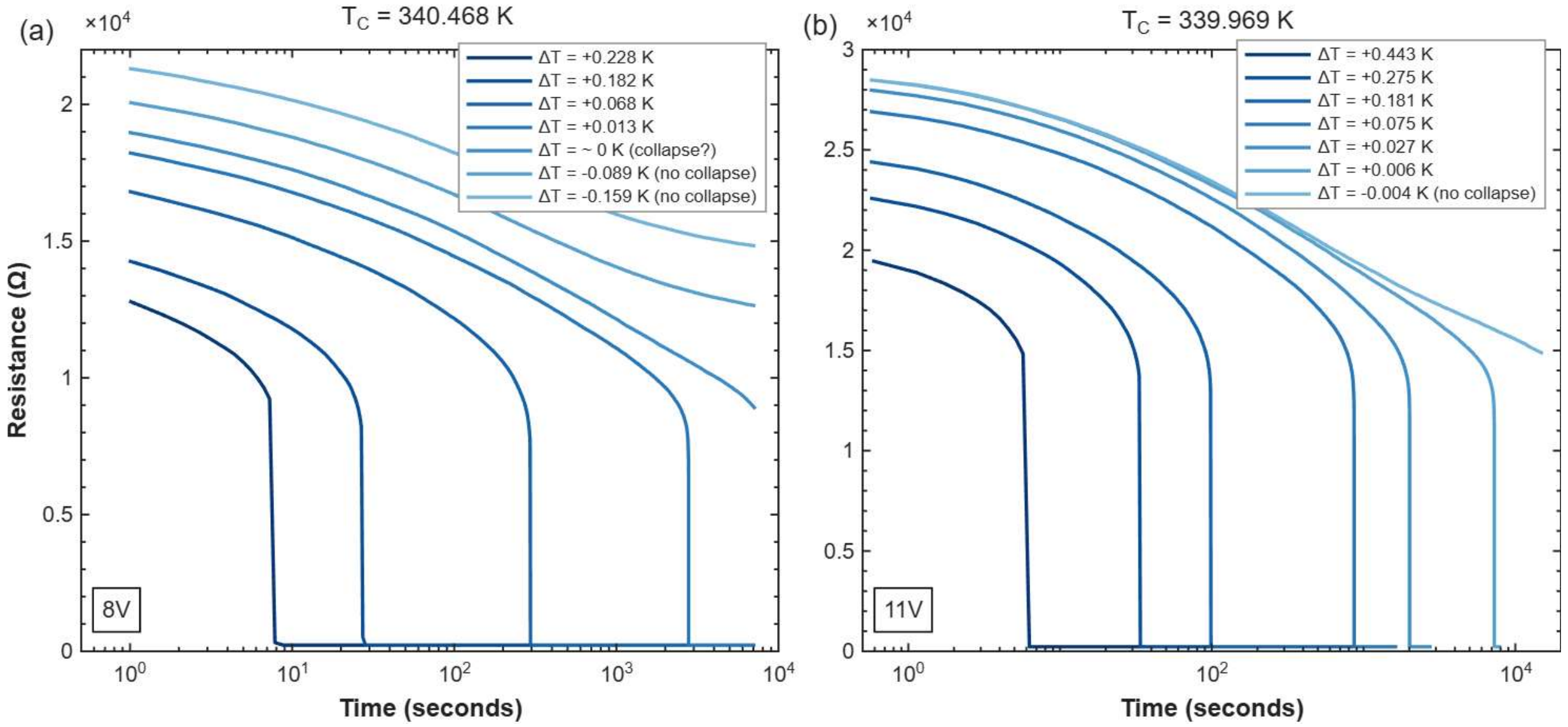

FIG. S4: Log-scaled time evolution of network resistance $R(t)$ at constant applied voltages of **(a)** 8 V and **(b)** 11 V across several stage temperatures near $T_c$. At temperatures above $T_c$, the device exhibits a characteristic incubation delay prior to rapid avalanche switching. As the temperature approaches $T_c$, the delay time grows by several orders of magnitude. For $T < T_c$, complete switching is not reached, and for $T \sim T_C$ the switching did not occur within the measurement window ($t \leq 10^4$ s).

### S4. Additional plateau measurements

Below are additional demonstrations of the network's temporal resistance evolution, for dwells at 8V and 11 V (Fig. S4). As in Fig. 3 of the main article, we can see here the plateau, the latency prior to the abrupt resistance drop, $\tau_p$, increases as the temperature approaches criticality (the transition temperature). Note that, unlike Fig. 3(a) of the main text, which uses a linear time axis, Fig. S4 uses a logarithmic time axis.

## S5. From the resistance record to the branching factor

#### What the record counts

Under a fixed bias voltage, the resistance of the $VO_2$ network decays over $10^2$–$10^4$ s in a staircase of discrete steps [Fig. S5.1(a), inset]. Each step is a region (or regions) of the film switching from the insulating to the metallic phase. The analysis treats the record as a point process of such switching events and asks a single question: is the rate of events accelerating or decaying?

We work with the conductance $G = 1/R$ rather than with $R$, because $G$ is the extensive counter. The conducting regions act in parallel, so to leading order each newly metallic region adds a roughly constant increment to the total, whereas the corresponding resistance step, $\Delta R = R^2 \Delta G$, shrinks with $R^2$ by a different factor (4.6–6.3 across the plateau in the three collapsing runs from the main article). A symmetric two-dimensional effective-medium treatment (Sec. S3) [1,2,2] places the metallic area fraction at $f = 0.39$ at the start of the plateau, rising to $f = 0.47$ at the collapse, i.e. still below the two-dimensional threshold $f_c = ½$ [3], and $dG/df$ varies smoothly over that range. This is the mean-field assumption we use.

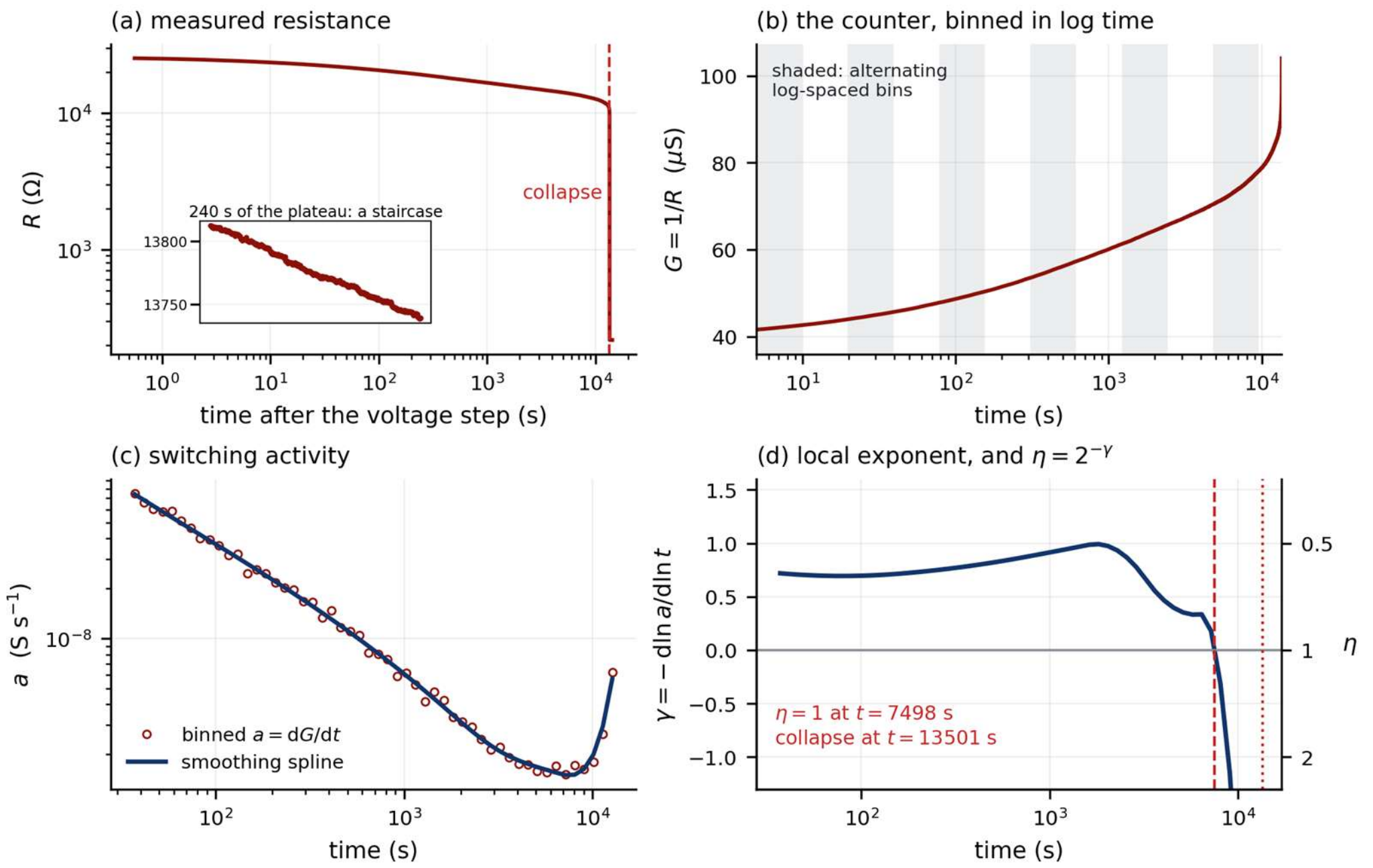

**FIG. S5.1.** The analysis, stage by stage, on one collapsing run (10 V, 339.94 K). (a) The measured resistance; the inset resolves 240 s of the plateau into discrete switching steps. (b) The conductance $G = 1/R$, with alternating logarithmic bins shaded. (c) The activity $a = \mathrm{d}G/\mathrm{d}t$ from the binned slopes (circles) and the smoothing spline (blue). (d) The local exponent $\gamma = -\mathrm{d}\ln(a)/\mathrm{d}\ln(t)$ and the branching factor $\eta = 2^{-\gamma}$ (right axis). The dashed line marks the crossing $\eta = 1$ and the dotted line marks the collapse, which occurs 1.8 times later.

### The switching activity

The switching activity is defined:

$$a(t) \equiv dG/dt\,, \quad \text{(S5.1)}$$

obtained as an unweighted least-squares slope of $G(t)$ inside logarithmically spaced time bins (between 50 and 100 bins were used, starting from 5 s and going to the end of the plateau). Logarithmic spacing is imposed by the data since the plateau spans between two and three and a half decades of time, so a uniform grid over-resolves late times and under-resolves early ones. A cubic smoothing spline through $ln\,(a)$ vs. $ln\,(t)$ supplies the differentiable curve used below, and the number of bins and the spline smoothing are the free parameters of the procedure. Figure S5.1 shows the four stages.

**The branching factor**

From the activity we form the local exponent and define a branching factor estimator (motivated by ref. [4]):

$$\gamma(t) = -d\ln(a)/d\ln(t), \quad \eta(t) = 2^{-\gamma(t)}. \quad \text{(S5.2)}$$

η is the factor by which the switching rate changes per doubling of the elapsed time. This means that $\gamma > 0$ ($\eta < 1$) is a decaying cascade and $\gamma < 0$ ($\eta > 1$) is an accelerating one, exactly as expected for a branching factor. In general, a branching process is bookkept in generations of some duration τ, with σ the mean number of elements switched by one switched element [5,6]. Its activity obeys $a(t + \tau) = \sigma\, a(t)$, so $d\ln(a)/dt = \ln(\sigma)/\tau$, and converting the derivative to the logarithmic clock with $d\ln(t) = dt/t$ gives

$$\gamma(t) = -t\ln(\sigma)/\tau. \quad \text{(S5.3)}$$

Hence, for any finite generation time τ and any $t > 0$, $\boldsymbol{\gamma = 0 \Leftrightarrow \sigma = 1}$. The zero of the measured exponent is the critical point of the underlying branching process, whatever τ is and whatever functional form $a(t)$ has - both conditions indicate that the switching rate has stopped changing. Two limitations follow from the same equation and we state them explicitly. (i) Away from criticality γ and σ are different numbers, related by $\sigma = \exp(-\gamma\tau/t)$, and since τ is not accessible in our experiment, we use η as a *criterion* for criticality and not as a measurement of the branching ratio away from it . (ii) The base 2 in Eq. (S5.2) is a convention that makes the critical value read 1 with one doubling of the clock as the unit; any other base rescales γ without moving its zero.

**Why the increment-ratio estimator cannot be used here**

The branching factor has been previously estimated from the ratio of successive increments of the order parameter, $\eta(t; \Delta t) = [X(t+\Delta t) - X(t)] / [X(t) - X(t-\Delta t)]$ [6,7], which in terms of the activity is simply $a(t+\Delta t)/a(t)$. In our measurements, when the dwell is long, $a(t)$ and $a(t+\Delta t)$ are both very small, so measurement noise makes this estimator very noisy and strongly dependent on the choice of $\Delta t$, which shifts the time at which $\eta = 1$, or eliminates the crossing altogether [8,9], because, unlike in other cases, the activity here decays over several decades in time.

**Validations**

**A cascade of known branching ratio.** We build an activity generation by generation from a prescribed $\sigma(t) = \exp(-\gamma_0(t)\,\tau/t)$, with $\gamma_0$ passing through zero at $t = 300$, and then apply to it the estimator used on the data. It recovers the imposed exponent across the whole range and places $\eta = 1$ at $t = 305$, within 1.8 % of the true critical time; at the true critical time $\eta = 0.9984$ [Fig. S5.2(a)].

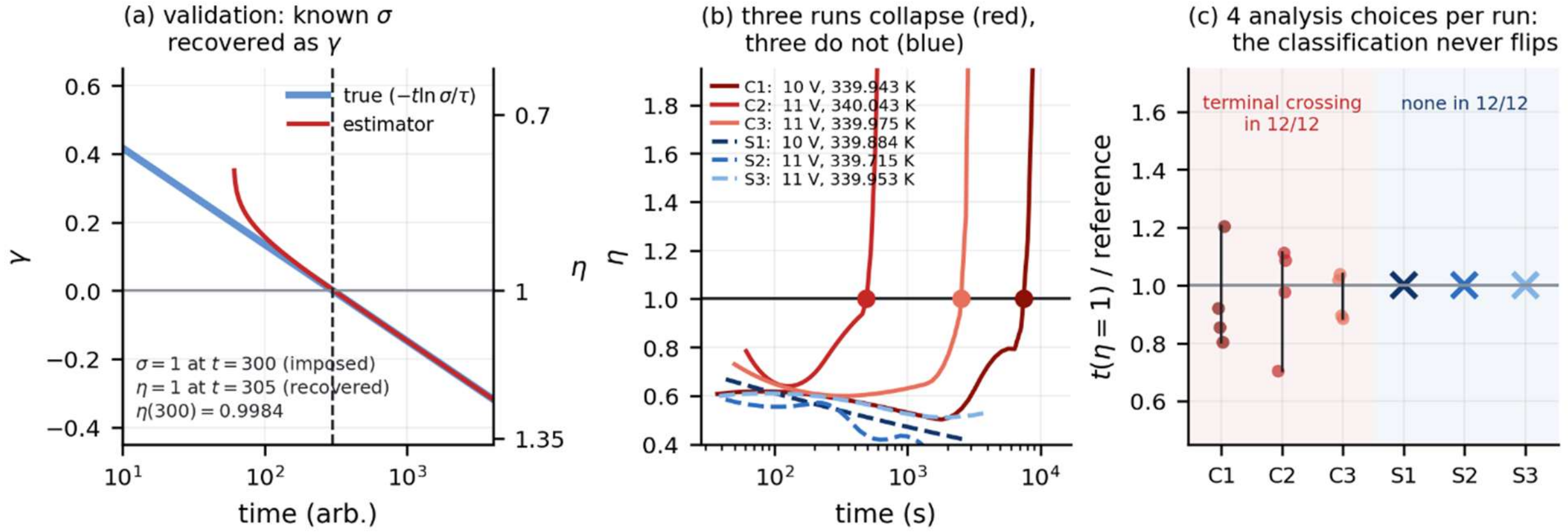

**FIG. S5.2.** Validation and robustness. (a) The estimator applied to a synthetic cascade built from a known branching ratio: it recovers the imposed exponent (blue) and locates η = 1 within 1.8 % of the imposed critical time. (b) η(t) for six runs; C1–C3 (red, solid) collapse, S1–S3 (blue, dashed) do not. Filled circles mark the crossings. (c) The crossing time under four analysis settings per run (two spline widths × two bin counts), normalized to the reference value of each run. Crosses denote runs for which no terminal crossing exists at any setting. **The classification never flips; the crossing time carries a 16–42 % systematic.**

**Collapsing and surviving runs are separated.** Figure S5.2(b) shows six runs at two bias values and neighboring stage temperatures, three ending in an abrupt collapse and three that do not. All three collapsing runs cross η = 1 and never return; no survivor does. We call a crossing *terminal* when γ does not return positive between the crossing and the end of the record — the acceleration, once begun, is not undone.

**Sensitivity to the analysis choices.** Repeating the whole procedure over a grid of two spline smoothing widths (s = 0.002 and 0.008) × two bin counts (70 and 100), i.e. four settings per run, is shown in Fig. S5.2(c). Each of the three collapsing runs produces a terminal crossing at every one of its four settings — twelve out of twelve — and none of the twelve settings applied to the surviving runs produces one. The classification is therefore not an artifact of the analysis choices. What is *not* sharp is the crossing time, which moves by 16–42 % peak-to-peak across the grid, and we quote *t*(η = 1) with that systematic attached. Interestingly, the ratio between the collapse time and the time of η = 1 is similar for the three runs: 1.80, 1.77 and 1.68 for the three runs. What matters is that the branching factor reaches unity well before the collapse. That separation, and not the precise value of the crossing time, is the claim the method supports.